\documentclass[10pt,twocolumn,aps,pra,floatfix]{revtex4-1}
\usepackage{todonotes}
\usepackage{graphicx}
\usepackage{import}
\usepackage{dcolumn}
\usepackage{bm}
\usepackage{hyperref} 
\usepackage[T1]{fontenc}
\usepackage{siunitx}
\DeclareSIUnit\pixel{px}
\DeclareSIUnit\fps{fps}
\DeclareSIUnit\counts{counts}
\definecolor{FluorGreen}{RGB}{141, 255, 141}
\definecolor{IR_Yellow}{rgb}{1.000, 0.784, 0.000}
\definecolor{ForestGreen}{RGB}{24, 170, 24}
\newcommand{\mdp}{$190\,\pm\,30$\,\si{\femto\watt\per\second\tothe{1/2}}}

\begin{document}
%\linenumbers
\title{Ultra-high-speed Terahertz Imaging using Atomic Vapour}

\author{Lucy A. Downes*$^1$}
\author{Andrew R. MacKellar$^1$}
\author{Daniel J. Whiting$^1$}
\author{Cyril Bourgenot$^2$}
\author{Charles S. Adams$^1$}
\author{Kevin J. Weatherill$^1$}
\affiliation{$^1$Joint Quantum Centre (Durham-Newcastle), Department of Physics, Durham University, South Road, Durham. DH1 3LE, UK.}
\affiliation{$^2$Centre for Advanced Instrumentation, Department of Physics, Durham University, NETPark Research Institute, Joseph Swan Road, Sedgefield, TS21 3FB, UK.}

\date{\today}

\pacs{Valid PACS appear here}

\maketitle

{\bf \noindent Terahertz (THz) technologies, generally defined as operating in the 0.1--10$\,$THz range, bridge the gap between electronic and photonic devices \cite{Dhillon17, Mittleman:18}. 
Because THz radiation passes readily through materials such as plastics, paper and cloth it can be employed in non-destructive testing \cite{Wietzke07},
and as it is non-ionising it is considered safe for security \cite{Federici05} and biomedical applications \cite{Woodward03}. 
There is significant demand for high speed THz imaging across a range of applications but, despite ongoing efforts, fast full-field imaging remains an unfulfilled goal.
Here we demonstrate a THz imaging system based upon efficient THz-to-optical conversion in atomic vapour \cite{Wade17}, where full-field images can be collected at ultra-high speeds using conventional optical camera technology. 
For a 0.55~THz field we show an effective 1\,\si{\centi\metre}$^2$ sensor with near diffraction-limited spatial resolution and a minimum detectable power of $\bm{190\,\pm\,30}$\,\si{\femto\watt\second}$\bm{{}^{-1/2}}$\, per $\bm{40\times40}$\,\si{\micro\metre} pixel capable of video capture at 3000 frames per second. 
This combination of speed and sensitivity represents a step change in the state of the art of THz imaging, and will likely lead to its uptake in wider industrial settings. With further improvements we expect that even higher frame rates of up to 1~MHz would be possible.
}

Numerous THz imaging technologies are currently available, each with its own merits and challenges. 
Single point detectors can be used to collect high-resolution images pixel by pixel \cite{Chen2014, Stantchev2016} but have yet to demonstrate real-time capabilities.
For example, techniques employing thermal detectors such as Golay cells and bolometers are limited to response times on the order of 100\,\si{\milli\second} per pixel \cite{Dean2014}. 
While some techniques such as self-mixing in quantum cascade lasers (QCLs) can capture pixel data at significantly higher rates \cite{Dean08}, they are limited by the speed at which they can raster between pixels \cite{Rothbart2013}.
Full-field THz imaging techniques, in which a complete image is collected in a single shot, eliminate the need to scan between individual pixels. These schemes typically consist of arrays of point detectors such as microbolometers \cite{Lee06}, field-effect transistors (FETs) \cite{Qin17} or carbon nanotubes \cite{Suzuki2016}. Such arrays can operate up to video frame-rates of around 60\,\si{\hertz} but their sensitivities are limited and scaling to large pixel numbers can be challenging \cite{Simoens_Meilhan:2013}. 
More sensitive arrays such as those based on superconductors \cite{Ariyoshi06} suffer similar scaling problems and rely on cryogenics to overcome thermal noise, making them cumbersome, expensive and complex to operate.
Another promising approach to full-field THz imaging is frequency up-conversion, where a nonlinear optical material such as diamond \cite{Clerici:13}
is used to convert incident THz photons to more easily-detectable wavelengths allowing images to be collected using standard optical or infra-red cameras. 
The nonlinear conversion efficiency is typically low and thus averaging over many high intensity THz pulses is required to obtain usable optical signal levels, resulting in limited frame rates.  Despite the high demand, a room-temperature THz imaging system combining both high speed and sensitivity has yet to be realised \cite{Chen2014, Dhillon17, Mittleman:18}.

%%%%%%%%%%%%%%%%%%%%%%%%%%%%%%%%
%\section{Experimental Setup}
%%%%%%%%%%%%%%%%%%%%%%%%%%%%%%%%
\begin{figure*}[ht]
\centering
\def\svgwidth{1.0\linewidth}
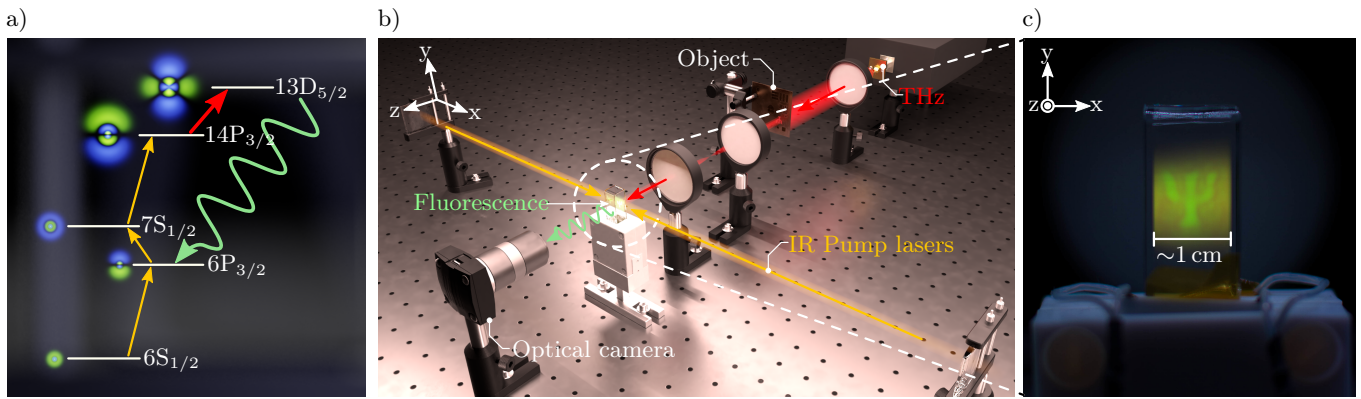
\caption{\textbf{THz imaging using atomic vapour.} (a) The internal energy structure of caesium, highlighting the excitation and principal decay pathways used in this work. The orange arrows correspond to laser excitations, the red arrow to the excitation driven by the THz field and the green arrow to the dominant fluorescence decay. (b) A diagram of the experimental set-up described in this work. Infrared pump lasers form a sheet of excited caesium atoms contained within a quartz cell. The THz field passes through an object which is imaged onto the atoms using PTFE lenses. Interaction of the THz field with the excited atoms leads to the emission of green fluorescence which is captured using an optical camera. (c) An example of a true-colour THz image of a `$\Psi$'-shaped aperture formed within the vapour cell.}
\end{figure*}

%%%%%%%%%%%%%%%%%%%%%%%%%%%%%%%%%%%%%%%%%%%%%%%%%%%%%%%%%%%%%%%%%%%%%%%%%%%%%%%%%%%%%%%%%%%%%%%%

In this work we demonstrate and characterise a high-speed THz imaging system based on efficient THz-to-optical conversion in a room temperature atomic vapour \cite{Wade17}. The principle of operation is to use resonant transitions in atoms to convert difficult-to-detect THz radiation into easy-to-detect optical photons. 
We use atomic caesium due to its high vapour pressure and easily accessible transition wavelengths. 
The atomic structure involved in the process is shown in Fig.~1a. The vapour is prepared in the 14P$_{3/2}$ state using three infrared (IR) lasers from where a resonant THz field transfers population into the nearby 13D$_{5/2}$ state. 
From here the population can decay back to the intermediate 6P$_{3/2}$ state by emitting green (535\,\si{\nano\metre}) fluorescence, which can subsequently be detected using an optical sensor. 
The versatility of this approach means that any optical camera can be employed, and can easily be substituted into the system.
A schematic of the imaging setup is shown in Fig.~1b.
The caesium vapour is contained within a quartz cell and prepared using co-axial IR laser beams shaped such that a 100\,\si{\micro\metre} thick sheet of excited atoms is formed in the $xy$ plane.
The low-power continuous-wave THz field (up to 19\,\si{\micro\watt} at 0.55\,\si{\tera\hertz}) propagates in the $z$ direction, perpendicular to the plane of the vapour. Along this axis a 1:1 transmission imaging system, comprised of PTFE aspheric lenses, is used to project a THz image onto the sheet of excited atoms (Fig.~1b). 
The fluorescence emitted by atoms in the region of overlap between the lasers and the THz beam is re-imaged onto an optical camera, providing an image of the incident THz field in a single shot. To isolate fluorescence from the THz-coupled state we employ a narrowband optical filter to minimise background fluorescence reaching the camera.
Three different optical cameras (A, B, C) were used in this work, details of which can be found in Methods. 
An example of a true-colour, unfiltered THz image formed within the vapour cell is shown in Fig.~1c. The `$\Psi$' shape, clearly visible in green fluorescence, is created by placing a metal mask (see Fig.~2) in the object plane.

%%%%%%%%%%%%%%%%%%%%%%%%%%%%%%%%%%%%%%%
%\section{Spatial Resolution}
%%%%%%%%%%%%%%%%%%%%%%%%%%%%%%%%%%%%%%%
\begin{figure*}[h]
\centering
\def\svgwidth{1.0\linewidth}
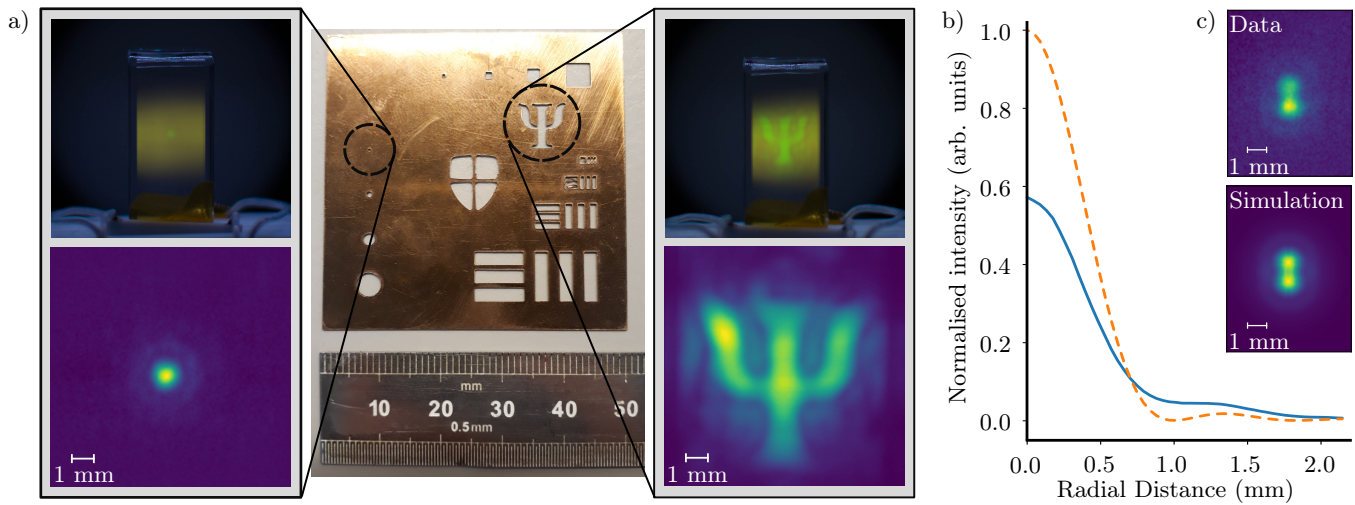
\caption{\textbf{Demonstration of spatial resolution.} (a) A metal mask (centre) is placed in the object plane of the system. To the left and right are true colour images taken with camera A (above) and false colour images taken with camera B (below) for a 0.50\,\si{\milli\metre} diameter pinhole (left) and a `$\Psi$'-shaped aperture (right). (b) Radially averaged intensity profiles of the measured PSF (solid blue line) and the simulated ideal PSF (dashed orange line). Both PSFs were scaled such that the total power in each image was equal. The ratio of peak heights gives a Strehl ratio of 0.57. (c) Real (top) and simulated (bottom) images of two 0.50\,\si{\milli\metre} diameter pinholes separated by 1.00\,\si{\milli\metre}.} \label{fig:2}
\end{figure*}

To test the spatial resolution of the system we image a metallic `test card' comprising apertures of varying sizes and shapes (see Fig.~2a). 
Images of a 0.50\,\si{\milli\metre} diameter pinhole and `$\Psi$'-shaped aperture are shown both as true-colour unprocessed photographs taken with camera A and as false colour filtered images taken with camera B.
By considering the image of the 0.50\,\si{\milli\metre} diameter pinhole to be the point spread function (PSF) of the system we are able to compare the performance to that of an ideal imaging system.
We normalise both the real and ideal unaberrated PSF such that the total power in each image is equal.
We then perform a radial average about the point of peak intensity, the results of which are plotted in Fig.~2b.
By calculating the ratio of the maximum peak heights we extract a Strehl ratio for our system of 0.57, indicative of a moderately aberrant system.
We posit that these aberrations arise from the THz lenses and simple design of the imaging system used.
By considering the Rayleigh criterion we expect that our system will be able to resolve two point sources separated by 1.00\,\si{\milli\metre}. 
We demonstrate this ability by imaging two 0.50\,\si{\milli\metre} diameter pinholes separated by 1.00\,\si{\milli\metre} (Fig.\,2c). 
The resulting image (top) closely resembles the simulated image (bottom) in which the two distinct apertures are clearly resolved, providing an upper bound of 1.00\,\si{\milli\metre} on the system's spatial resolution.
We conclude that the system achieves near-diffraction-limited resolution and is currently limited by aberrations in the THz optics.

%%%%%%%%%%%%%%%%%%%%%%%%%%%%%%%%%%%%%%
%\section{Temporal Resolution}
%%%%%%%%%%%%%%%%%%%%%%%%%%%%%%%%%%%%%%
To illustrate the high speed capabilities of this technique we demonstrate THz imaging of dynamical processes at frame rates up to 3\,\si{\kilo\hertz}, two orders of magnitude faster than the current state of the art \cite{Fan:15}.
Here, we use camera C to capture videos of a water droplet in free fall at a frame rate of 500\,\si{\hertz} and a rotating optical chopper wheel at 3\,\si{\kilo\hertz}. 
Frames from each video are presented in Fig.~3 (full video files are available in the online Supplementary Material).
Although the frames in Fig.~3b were taken at a frame rate of 500\,\si{\hertz}, only every fourth frame is shown to illustrate the changing shape of the water drop in free fall captured in the video. 
To improve image clarity we perform post-processing on the data,
further details of which can be found in Methods/Supplementary Material. 

\begin{figure*}
\centering
\def\svgwidth{1.0\linewidth}
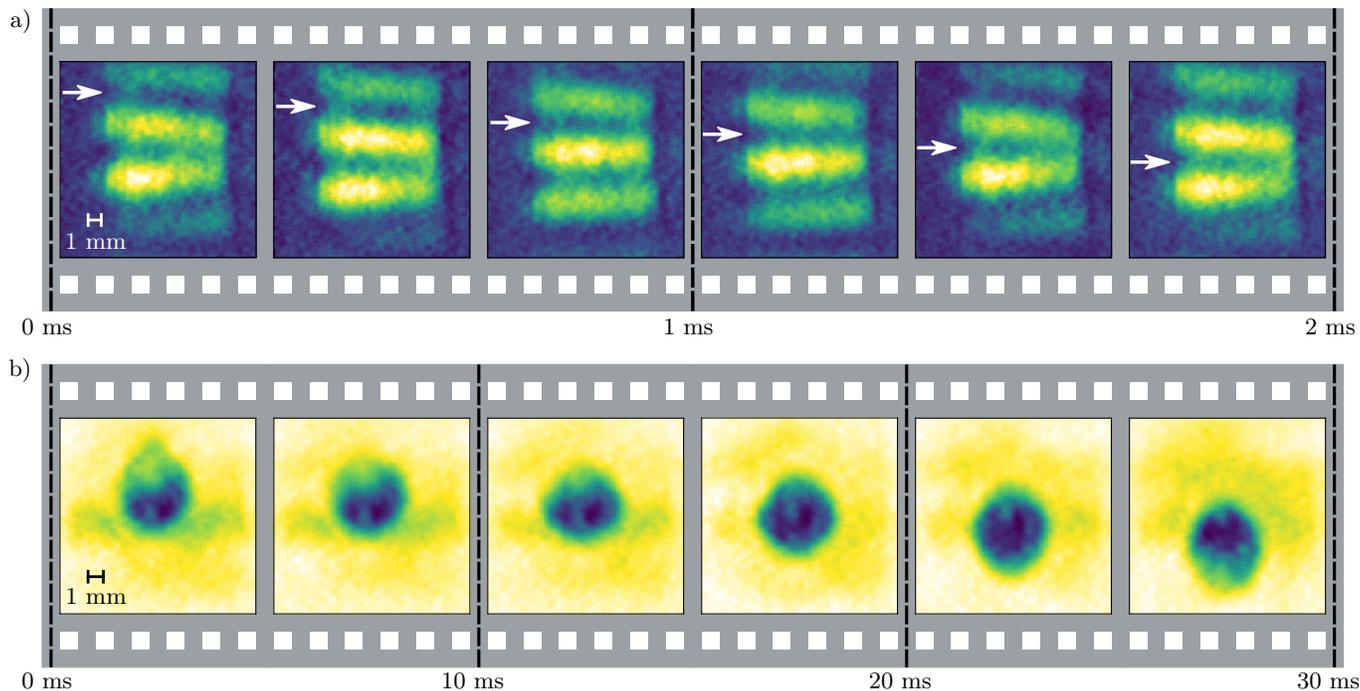
\caption{\textbf{Ultra-high-speed THz video.} (a) Subsequent frames from a THz video of an optical chopper wheel rotating at 700\,rpm, imaged at a frame rate of 3\,\si{\kilo\hertz}. The white arrow is added to highlight the movement of one spoke of the wheel between frames. (b) A significantly slower frame rate of 500\,\si{\hertz} is sufficient to capture the dynamics of a water droplet in free fall shortly after being released from a burette, every fourth frame of which is shown here. Full video files are available online.}
\end{figure*}

%%%%%%%%%%%%%%%%%%%%%%%%%%%%%%%%
%\section{Sensitivity}
%%%%%%%%%%%%%%%%%%%%%%%%%%%%%%%%
\begin{figure}
\centering
\def\svgwidth{1.0\linewidth}
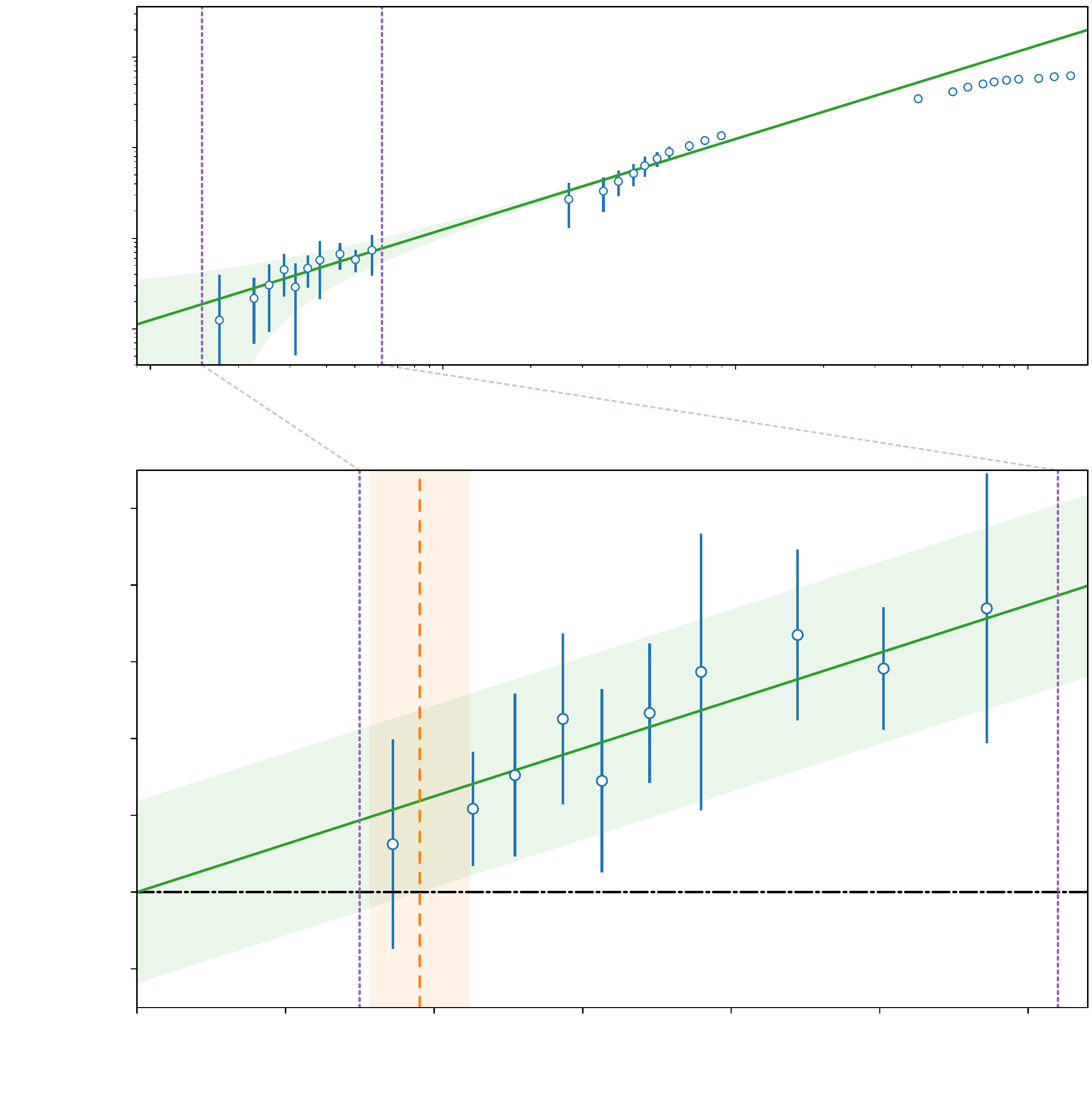
\caption{\textbf{Minimum Detectable Power.} (a) The fluorescence signal $\rm{THz_{on}} - \rm{THz_{off}}$ (open blue circles) for varying THz powers for a total integration time of 1\,\si{\second}. The linear trendline is extrapolated from that in (b) to highlight the saturation point at THz powers above 20\,\si{\pico\watt\per\pixel}. The vertical dashed lines highlight the range of the data used in (b). (b) Plotting the measured fluorescence signal (open blue circles), we can map the linear response of the system (green line) plus its associated uncertainty (green shaded region) to the point at which the fluorescence signal is no longer reliably detectable (orange dotted line), at $190\,\pm\,30$\,\si{\femto\watt\per\pixel} for a 1\,\si{\second} integration time.} \label{fig:4}
\end{figure}

We characterise the detector sensitivity by measuring the minimum detectable power (MDP); 
the minimum THz power at which the resulting optical signal is reliably detectable above the noise. Camera B was used to record a series of images, both with and without the incident THz field for varying THz power.
We define the signal as the pixel value resulting from fluorescence in the presence of the THz field minus that from background fluorescence in the absence of the THz field, $\rm{THz_{on}} - \rm{THz_{off}}$.
For a typical $40\times40$~\si{\micro\metre} pixel close to the centre of the image, this background-subtracted signal and its uncertainty is plotted against the incident THz power in Fig.~4a \cite{Hughes10}.
We find the system responds linearly for THz powers up to around 20\,\si{\pico\watt\per\pixel}. 
Above this the system experiences saturation and we observe a smaller increase in signal for a given increase in THz power, seen as the deviation from the linear trend at high THz powers.
We determine the MDP by considering the signal in the region of lowest incident THz power shown in Fig.~4b. From this we find the minimum THz power at which the signal rises above its associated uncertainty to be  $190\,\pm\,30$\,\si{\femto\watt} for a 1\,\si{\second} exposure, as indicated by the orange vertical line in Fig.~4b.
At exposure times of over 0.5\,\si{\second} the fluorescence signal saturates the camera CCD, so to obtain an integration time of 1 second we average over 5 frames, each with an exposure of 200\,\si{\milli\second}. 
At this exposure time we are working within the shot-noise limited regime of the camera where the recorded pixel value scales linearly with exposure time, and the uncertainty is proportional to the square root of the pixel value. 
In this way the MDP of our system scales inversely with the square root of the total integrated exposure time used, resulting in an MDP of \mdp\, per $40\times40$\,\si{\micro\metre} pixel. 
This represents an improvement of over two orders of magnitude on other room temperature imaging systems \cite{Simoens_Meilhan:2013} and is the reason that such high frame rates are possible.

%%%%%%%%%%%%%%%%%%%%%%%%%%%%%
%\section{Discussion}
%%%%%%%%%%%%%%%%%%%%%%%%%%%%%
Despite already demonstrating significant improvements in speed and sensitivity over other THz imaging systems, many relatively simple adjustments could be made to improve performance further.
For example, image quality could be improved by reducing the interference patterns caused by reflections of the THz field within the cell through making the cell thinner (<200\,\si{\micro\metre}) and adding an anti-reflection coating.
Furthermore, the image resolution could be increased by using a more sophisticated THz lens system, or by imaging using higher THz frequencies through the choice of an appropriate atomic transition from the wide range available \cite{Wade17}.
The THz sensor area could be extended by using a larger vapour cell to enable formation of a larger sheet of excited atoms, however increased laser powers would be required to maintain laser beam intensity over a larger area. 
We further note that the response time of this system is ultimately limited by the lifetime of the atomic state used, which here is 0.80\,\si{\micro\second} \cite{Sibalic17}.
Our full-field approach enables the system to be used for capturing high-speed THz video, potentially up to \si{\mega\hertz} frame rates with a suitable optical camera.
As camera C is not designed for low-light applications, we are limited in this work to frame rates less than 4\,\si{\kilo\hertz}.

The technique presented here adds to a catalogue of ongoing work within the `Quantum Technologies' landscape \cite{Schleich16} where atoms are used as sensors. This includes recent demonstrations of atom-based EM-field sensing \cite{Fan:14,Holloway:14}, magnetic field sensing \cite{Budker2007, Dietsche2019} and gravimetry \cite{Hu2013,Kritsotakis18}. For these and other applications, there is an ongoing effort to miniaturise and simplify the required laser technology and associated electronic hardware. Since our THz imaging system uses the same underlying techniques it will benefit from the drive to bring atom-based technologies to market.

Whilst we highlight simple modifications which could be made to improve the performance of this system, the figures of merit reported here have enabled THz imaging of dynamical processes at rates not previously possible by any other method. We predict that the versatility and sensitivity of this atom-based THz imaging technique will produce a disruptive impact on fields as diverse as biological imaging and production-line quality control \cite{Jansen:10, Jepsen2011, Mittleman:18}.

{\bf Acknowledgements}\\
We thank Del Atkinson, Paul Dean, Ifan Hughes, Matt Jones, Marco Peccianti and Chris Wade for stimulating discussions. We thank Mike Tarbutt, Ray Sharples, Andrew Gallant and Claudio Balocco for the loan of equipment.
This work is supported by UK EPSRC Grants EP/S015973/1, EP/R002061/1, EP/M014398/1, EU.H2020.macQsimal ID:820393 and M Squared lasers.

{\bf Author Contributions}\\
KJW and CSA conceived the idea of the project. LD, ARM and DJW performed the measurements. CB manufactured the PTFE optics. LD, DJW and ARM performed the image analysis. All authors contributed to the writing of the manuscript. KJW managed the project.
% \bibliography{thzbib.bib}
% \bibliographystyle{pfg-physrev3}

\clearpage
\noindent{\bf Methods}
\vspace{0.3cm}

\textbf{Light sheet} --- The caesium vapour is contained in a cuboidal quartz cell with an optical path length of 1\,\si{\centi\metre}, which is heated to a temperature of $\sim$60\,\si{\celsius} using resistive metal-ceramic heaters. The vapour is excited to the 14P$_{3/2}$ state using a three-step excitation process~\cite{Carr12b}. The probe laser (852\,\si{\nano\metre}) excites atoms to the 6P$_{3/2}$ state, and the coupling laser (1470\,\si{\nano\metre}) takes atoms from the 6P$_{3/2}$ state to the 7S$_{1/2}$ state. Both probe and coupling laser frequencies are stabilised to their respective atomic resonances using polarisation spectroscopy~\cite{Carr:12}. The Rydberg laser (843\,\si{\nano\metre}) is tuned to the 7S$_{1/2}\rightarrow 14$P$_{3/2}$ transition, and is left free running. All three laser beams propagate co-axially and are shaped to form a light sheet of approximately 100\,\si{\micro\metre} in width and 10\,\si{\milli\metre} in height at the position of the vapour cell. The Rydberg laser propagates in the opposite sense to the probe and coupling beams, minimising the 3-photon Doppler shift due to atomic motion through the optical fields. The beam powers of the probe, coupling and Rydberg lasers are approximately 5\,\si{\milli\watt}, 20\,\si{\milli\watt} and 200\,\si{\milli\watt} respectively.

\textbf{Atomic States} --- The Rydberg states used in this work (14P$_{3/2}$ and 13D$_{5/2}$) were chosen to produce the clearest images. This was a balance between transferring the maximum number of atoms to the final Rydberg state while having a clear contrast between the signal and background fluorescence. The selected transition has a large dipole matrix element of $79.77\,a_0 e$ meaning the atoms couple strongly to the THz field, while having one dominant decay pathway leading to maximal signal photons. 
The characteristics of the Rydberg states were calculated using the Alkali Rydberg Calculator (ARC) package in Python \cite{Sibalic17}.

\textbf{THz system} --- The free-space continuous-wave Terahertz beam (0.55\,\si{\tera\hertz}) is generated using a amplifier multiplier chain (AMC), manufactured by \emph{Virginia Diodes Inc.}, which is seeded by a microwave signal generator. 
The THz beam is primarily linearly polarised in the vertical ($y$) direction and is tuned to the frequency of the 14P$_{3/2}\rightarrow 13$D$_{5/2}$ transition.
The THz power is controlled coarsely by the addition of 50\,\si{\milli\metre} thick Nylon attenuators, with fine control provided by an internal voltage controlled attenuator. 
A photo-accoustic power meter (TK TeraHertz Absolute Power Meter System) was used to measure the maximum output power of the THz source (19\,\si{\micro\watt} at 0.55THz) and to calibrate the the internal attenuator. 
The THz beam is launched into free space using a diagonal horn antenna, a 2" f = 75\,\si{\milli\metre} PTFE lens collimates the output, providing uniform illumination of the mask. 
Two custom 2" f = 75\,\si{\milli\metre} aspheric PTFE lenses (radius of curvature of 32.425\,\si{\milli\metre} and conic constant of -0.57244) form a conjugate image of the object in the plane of the atoms. 

\textbf{Cameras} --- In this work three different optical cameras were employed. The true colour images were taken using a commercial Nikon D5500 DSLR (Camera A) with an exposure time of 0.5\,\si{\second}, an F-stop of f/1.8 and an ISO of 100. 
An Andor iXon EMCCD camera (Camera B) was used for high resolution low-noise images, running without water-cooling to the EMCCD. An exposure time of 200\,\si{\milli\second} was used for this camera with the EM gain set to minimum. 
A Photron FASTCAM SA4 (Camera C) was used to record the high speed video at exposure times equal to the reciprocal of the frame rates stated. When using cameras B and C we use a simple 1:1 optical triplet lens to image the vapour, and narrowband filters [BrightLine Fluorescence Filters 535/6 \& 505/119] to remove unwanted IR scatter and background fluorescence. 

\textbf{Image processing} --- In all except the true-colour images, an image of the cell without the THz field was subtracted to remove any background fluorescence. 
The high resolution iXon images were captured directly using the PyAndor Python package, imported into Python, binned to form $4\times 4$ superpixels and smoothed using a uniform filter (\verb|scipy.ndimage.uniform_filter| with \verb|size = 3|).
The high-speed videos were captured using the Photron FASTCAM Viewer 3 software package.
The clarity of these images was improved by performing identical post-processing on each of the frames to remove the added camera noise. 
The images are binned ($8\times8$) and a spatial Fourier filter is applied to remove the grid-like artefact added by the camera. To do this we perform a 2D FFT and mask out the clear periodic structures corresponding to the grid noise. We then perform the inverse 2D FFT and the images are smoothed. 
Further details can be found in the Supplementary Material. The videos are comprised of a sequence of processed still frames and compiled using ffmpeg.

All data and codes are available online at DOI:10.15128/r16w924b81j

\bibliography{thzbib.bib}
\bibliographystyle{pfg-physrev3}

%%%%%%%%%%%%%%%%%%%%%%%%%%%%%%%%%%%%%%%%%%%%%%%
%% SUPPLEMENTARY INFO
%%%%%%%%%%%%%%%%%%%%%%%%%%%%%%%%%%%%%%%%%%%%%%%

\clearpage
\begin{center}
   {\Large \bf Supplementary Information} 
\end{center}

\section{Image Processing}

The images displayed in Fig.~3 have undergone post-processing to improve clarity and reduce noise. 
First an image in the absence of the THz field is subtracted to remove unwanted background fluorescence, then the resulting image is binned into $8\times8$ superpixels.
This reveals a periodic grid-like structure which is varies between frames and so cannot be removed by subtracting a blank field image.
To remove this grid while retaining as much of the image data as possible we perform a spatial Fourier filter to eliminate the unwanted structure.
To do this we perform a 2D FFT (\verb|numpy.fft.fft2|) on the binned image and manually mask out the features corresponding to the periodic grid structure by setting these regions equal to zero. 
We then perform the inverse 2D FFT (\verb|numpy.fft.ifft2|) to recover the filtered image.
To further improve image clarity we smooth the filtered image using a uniform filter (\verb|scipy.ndimage.uniform_filter|) with \verb|size = 3|.
Each frame taken with camera C was processed in this fashion, including the frames used to make the videos.

\begin{figure}[h]
\centering
\def\svgwidth{0.8\linewidth}
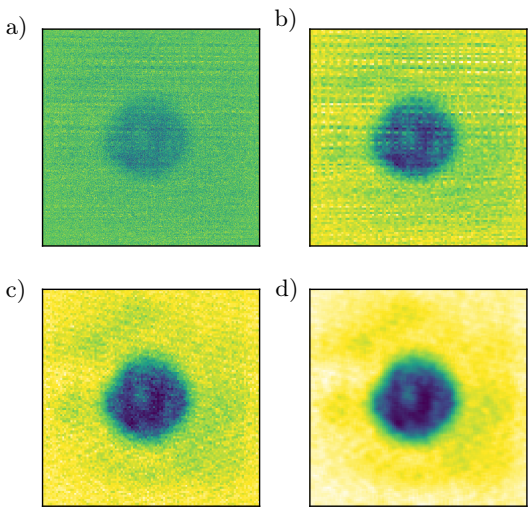
\caption{\textbf{Image Processing.} The sequence of steps used to process frames taken with camera C. The image used here is the same as that shown in the 4th frame of Fig.~3b. (a) The raw camera image with the background subtracted. (b) The resulting image after performing $8\times8$ binning. The grid-like structure is most apparent here. (c) The result of applying the spatial Fourier filter. The grid-like structure is mostly removed without distorting the image. (d) The final image after applying the uniform filter.}
\end{figure}

\section{Airy Pattern}

To determine the spatial resolution of the system we compared an image of a 0.50\,\si{\milli\metre} pinhole taken with our system to a diffraction-limited image.
Since the diameter of the pinhole is sub-wavelength, we assume that it can be treated as a point source and hence that the intensity $I$ at a radial distance $r$ from the centre will be given by an Airy pattern \cite{FtoF}, 

\begin{equation}
    I(a) = I_0\left(\frac{2J_1(a)}{a}\right)^2
\end{equation}
where
\begin{equation}
    a = \frac{\pi r}{\lambda N}.
\end{equation}
Here $\lambda$ is the wavelength of the imaging light, equal to 0.55\,\si{\milli\metre} and $N$ is the f-number of the imaging system.
The factor $I_0$ defines the maximum intensity at the centre of the image.
Based on the characteristics of the THz lenses, we calculate the f-number of the system to be 1.5 using
\begin{equation}
    N = \frac{f}{D},
\end{equation}
where $f$ is the focal length of the lens and $D$ is its effective aperture.

\section{Minimum Detectable Power Scaling}

When measuring the minimum detectable power (MDP) we stated that the value scaled as $\rm{s}^{-1/2}$. 
This comes about from the way in which the MDP is calculated, and the time scalings within the system.
In Fig.~4a we find the MDP as the point at which the fluorescence signal rises above zero by more than its associated uncertainty.
The fluorescence signal $F_{\rm{sig}} = \rm{THz_{on}} - \rm{THz_{off}}$ is linear with low incident THz powers $P_{\rm{THz}}$, and can be expressed as 
\begin{equation}
    F_{\rm{sig}} = n P_{\rm{THz}}
\end{equation}
where $n$ is the increase in signal per unit of THz power.
The location of the minimum detectable power is then found as the incident THz power for which 
\begin{equation}
    F_{\rm{sig}} - \alpha_{F_{\rm{sig}}} = 0
\end{equation}
where $\alpha_{F_{\rm{sig}}}$ is the uncertainty in the measure of the fluorescence signal.
Since we are working in the shot-noise limited region of the camera, the uncertainty on the measured fluorescence will be proportional to the square root of the measured value. 
Assuming that our signal is small compared to the pixel values in each frame such that $\rm{THz_{on}} \approx \rm{THz_{off}}$, the error in the signal can be expressed as
\begin{equation}
    \alpha_{F_{\rm{sig}}} \approx \sqrt{2}\sqrt{\rm{THz_{off}}}.
\end{equation}
Combining the above yields an expression for the MDP,
\begin{equation}
    \mathrm{MDP} = \frac{\sqrt{2}\sqrt{\rm{THz_{off}}}}{n}.
\end{equation}
where the value of $n$ can be found as the gradient in Fig.~4b.
It can then be seen that since both $\rm{THz_{off}}$ and $n$ depend linearly on the integration time used in the measurement, the value of $P_{\rm{min}}$ will have an $\rm{s}^{-1/2}$ dependency.
Ideally we would use a single 1\,\si{\second} exposure for these measurements, however at exposure times over 0.5\,\si{\second} the fluorescence saturates the camera CCD and the pixel values become unreliable.
To overcome this we average over 5 frames each with an exposure of 0.2\,\si{\second} to form a total integration time of 1\,\si{\second}.
As long as the exposure time per frame is within the shot-noise limited region of the camera and the read-noise is negligible, the measure of minimum detectable power will not change.
In the case of averaging over $N$ frames, the uncertainty on the fluorescence signal will be given by
\begin{equation}
    \alpha_{F_{\rm{sig}}} = \sqrt{\left(\frac{\sigma_{\rm{THz_{on}}}}{\sqrt{N-1}}\right)^2 + \left(\frac{\sigma_{\rm{THz_{off}}}}{\sqrt{N-1}}\right)^2}
\end{equation}
where $\sigma_{\rm{THz_{on, off}}}$ is the standard deviation of the measured fluorescence values \cite{Hughes10}.
Assuming Poissonian statistics for the pixel values, the standard deviation of the distribution that the measured values of fluorescence are drawn from is given by the square root of the mean value $\overline{\rm{THz}}_{\rm{on, off}}$.
Since $\overline{\rm{THz}}_{\rm{on, off}}$ is proportional to the exposure time per frame, an exposure that is a factor of $N$ shorter will reduce $\alpha_{F_{\rm{sig}}}$ by a factor of $N$.
Similarly the gradient $n$ will be reduced by a factor of $N$, and hence the measured MDP will be unchanged.
The uncertainty on the MDP is found by considering the errors in the two values used in its calculation; the gradient $n$ and the signal error $\alpha_{F_{\rm{sig}}}$. The error in the gradient $\alpha_{n}$ is given by the least-squares algorithm used to fit to the datapoints. The error in the mean signal error $\alpha_{\alpha_{F}}$ is estimated as the standard error in the mean across the 10 datapoints collected. 
Considering these uncertainties as uncorrelated, we find the error in the MDP, $\alpha_{\rm{MDP}}$, as
\begin{equation}
    \alpha_{\rm{MDP}} = \mathrm{MDP}\sqrt{\left(\frac{\alpha_n}{n}\right)^2 + \left(\frac{\alpha_{\alpha_F}}{\alpha_{F_{\rm{sig}}}}\right)^2}.
\end{equation}
The data shown in Fig.~4 is for a typical pixel in the centre of the image, with coordinates (125, 125).
Calculated MDP values vary between pixels as the actual incident THz power varies across the image. 

\section{THz Power per Pixel Calculations}

When calculating the minimum detectable power we need to know what THz power is incident on a single pixel. 
To do this we assume that the THz beam is a perfect Gaussian, and that all the lenses are positioned at their focal lengths.
The specification sheet for the THz horn \cite{VDi_Horn} gives a 3\,dB beamwidth of 10\,\si{\degree}. 
This is the angle between the points at which the intensity falls to half of its peak value, equal to the full width half maximum (FWHM).
In the optical configuration used, the beam shape at the first lens is the same as the beam shape at the position of the light sheet.
Since the lens has a focal length of 75\,\si{\milli\metre}, the beam will have a FWHM of 13.1\,\si{\milli\metre} at the position of this first lens, indicated by the grey dashed circle in Fig.~2.
To find the power incident on the light sheet we first find the power incident on the circle circumscribed around the area of the light sheet (white dashed circle, Fig.~2).
\begin{figure}
\centering
\def\svgwidth{0.8\linewidth}
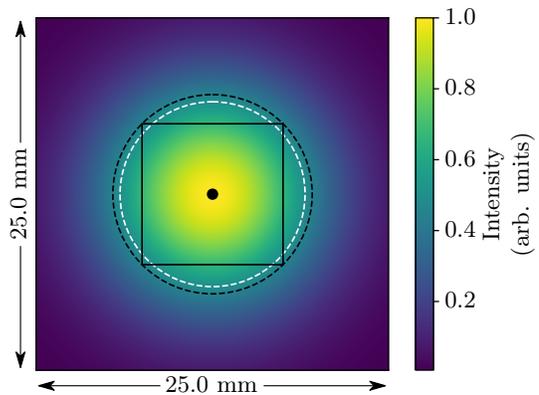
    \caption{An illustration of the geometry used in the power per pixel calculations. The colour represents the intensity of the THz beam at the position of the light sheet. The black square shows the approximate spatial extent of the light sheet, the black dashed line is the circumscribed circle. The white dashed line shows the location of the point at which the beam intensity is half of the peak value, the diameter of this circle is the FWHM.}
    \label{fig:my_label}
\end{figure}
Assuming the light sheet is a square with sides of 10.0\,\si{\milli\metre} in length (black box, Fig.~2), this circle has radius $r = 5\sqrt{2}$\,\si{\milli\metre}. 
For a Gaussian beam the fractional power $P$ transmitted through an aperture of radius $r$ is given by
\begin{equation}
    P = 1 - e^{\frac{-2r^2}{\omega^2}}
\end{equation}
where the beam waist $\omega$ is related to the FWHM through
\begin{equation}
    \omega = \frac{\rm{FWHM}}{\sqrt{2\rm{ln}2}}.
\end{equation}
Assuming that the beam is perfectly centred on the light sheet, we find that 55.6\% of the total power is incident within this circle. 
It then follows that 63.7\% of this incident power falls within the area of the light sheet.
We measure the maximum output power of the THz source to be 17\,\si{\micro\watt} at 0.55\,\si{\tera\hertz}, from which we find that a maximum of 6.02\,\si{\micro\watt} is incident over the area of the light sheet.
In the images used the light sheet covers $904\times904$ pixels, meaning that the power incident on the area covered by each pixel is 7.37\,\si{\pico\watt}.
For the image analysis the images were binned, increasing the maximum power per pixel by a factor of 16 to 118\,\si{\pico\watt}.
For coarse control over the incident THz power Nylon attenuators were placed in the beam path.
The transmission $T$ of the attenuator is given by
\begin{equation}
    T = e^{-\alpha l}
\end{equation}
where $\alpha$ is the absorption coefficient for the attenuator material and $l$ is its thickness.
Nylon has an absorption coefficient of $\alpha  = 5.5$\,\si{\per\centi\metre} at 0.55\,\si{\tera\hertz} \cite{Lockhart2014}, so each of the 0.50\,\si{\centi\metre} thick attenuators used reduces the THz power incident on the atoms by a factor of 15.6.

% \bibliography{thzbib.bib}
% \bibliographystyle{pfg-physrev3}

\end{document}